\documentclass[12pt,twocolumn]{iopart}

\usepackage{graphicx}% Include figure files
\usepackage{dcolumn}% Align table columns on decimal point
\usepackage{bm}% bold math
\usepackage{booktabs}
\usepackage{indentfirst}
\usepackage{color}
\begin{document}

\title{Magnetic excitations in the normal and nematic phases of iron pnictides}

\author{Hai-Yang Zhang$^{*}$, and Ning Xu$^{\dag}$}

\address{Department of Physics, Yancheng Institute of Technology, Yancheng 224051, China}
\ead{$^{*}$haiyangzhang03@163.com, $^{\dag}$xuning79530@126.com}
\begin{abstract}
In this paper, we study theoretically the behaviors of the magnetic excitations(MEs) in
the normal and nematic phases of iron pnictides. The normal state MEs exhibit commensurability to diamond and to square-like structure transition with the increase of energy. This structure transition persists in the spin and orbital scenarios of nematic phases, although the MEs show anisotropic behaviors due to the $C_{4}$ symmetry breaking induced by the nematic orders. The MEs exhibit distinct energy evolution behaviors between the spin and orbital scenarios of nematicity. For the spin-nematic scenario, the anisotropy of the MEs persists up to the high energy region. In contrast, for the orbital-nematic scenario, it reduces dramatically in the low energy region and is negligible in the high energy region. These distinct behaviors of the MEs are attributed to the different origins between the spin and orbital scenarios of nematic orders.
\end{abstract}

\section{The introduction}
The unconventional superconductivity found in
iron pnictides is of great interest in recent years. The
superconductivity arises when the carriers are doped into the parent compounds which
exhibit the collinear spin density wave(SDW) order below $T_{N}$.
Due to the proximity of the SDW and superconductivity phases, the magnetic fluctuations are proposed to
be responsible for the emergence of the unconventional superconductivity\cite{SCa,SCb,SCc,SCd}.
Experimentally, it was found that the SDW transition is generally accompanied by
a lattice distortion with the onset temperature $T_{s}$ above or coincident with $T_{N}$\cite{BFCA,BKCA,NFA}. It has been confirmed that the lattice structure-transition is attributed to the electronic nematicity which is characterized by the $C_{4}$ rotational symmetry breaking\cite{NMREV}. So far, it was proposed that the nematicity originates from the magnetic or orbital fluctuations\cite{NMREV,CFANG,CKXU,FERNANDES,ONARI}. Due to the tight relation between the SDW and the nematic phases, it is believed that there exists an intrinsic interplay between the nematicity and the magnetic fluctuations\cite{CFANG,CKXU,FERNANDES}. Thus, the study of the behaviors of the MEs and their relation to nematicity is highly desired as it may shed light on the mechanism of unconventional superconductivity.

So far, extensive inelastic neutron scattering(INS) experiments have been performed to investigate the behaviors of the magnetic excitations(MEs) in various iron-based compounds. It was found in the 122\cite{DAI122a,DAI122}, 111\cite{DAI111a,DAI111b} and 11\cite{Zhao11} materials that the MEs exhibit rich structure in the energy-momentum space. The low energy MEs are dominated by the commensurate peaks centered at $(\pi,0)$ and $(0,\pi)$. With the increase of energy, the MEs turn to be a diamond-like structure centered at $(\pi,\pi)$. A diamond to square structure transition occurs for the MEs with the further increase of energy. In addition, it was found that the MEs develop the anisotropic behaviors below a temperature $T^{*}$ which is basically coincident with the onset of the nematic fluctuations\cite{DAIprl2013,DAIscience2014,DAIprb2015}. The low energy MEs are dominated by the peaks around $(\pi,0)$ and $(0,\pi)$. However, they exhibit distinct intensity and temperature evolution between these two regions below $T^{*}$.

Theoretically, the behaviors of the MEs have been studied by some previous works. Especially, the effects of nematicity on the MEs are discussed within the orbital-nematic scenario. It was found\cite{Andersen2015} that the low energy MEs are dominated by peaks around $(\pi,0)$ and $(0,\pi)$. However, the peaks are different in intensity in the presence of the orbital order. This behavior of the MEs are qualitatively consistent with the INS data\cite{DAIscience2014,DAIprb2015}. The energy evolution of the MEs are detailed in Ref.\cite{Kreisel2015} over a wide energy region. Indeed, the theoretical results show that the MEs exhibit a structure transition with the increase of energy, similar to that observed by the INS experiments\cite{DAI122a,DAI122,DAI111a,DAI111b,Zhao11}. However, the origin of the structure transition has not been addressed. Besides, these previous works focus on the feedback effects of the orbital order on the behaviors of the MEs. However, the influences of the spin-nematic order on the behaviors of the MEs have not been studied so far. As we know, a detailed comparison of the MEs between the spin and orbital scenarios of nematicity is essential to understand the origin of nematicity\cite{NMREV} by using the INS technique.

In this paper, we study the behaviors of the MEs in the normal and nematic phases of iron-pnictides. For the normal state, the low energy MEs are dominated by the peaks centered at $(\pi,0)$ and $(0,\pi)$. With the increase of energy, the commensurate MEs give ways to a diamond-like structure with peaks centered at $(\pi\pm\delta,\pi)$ and $(\pi,\pi\pm\delta)$. With the further increase of energy, the MEs turn to be dominated by a square-like structure with peaks centered at $(\pi\pm\delta,\pi\pm\delta)$ and $(\pi\mp\delta,\pi\pm\delta)$. Thus, this energy evolution behaviors of the MEs are qualitatively consistent with the INS observations\cite{DAI122a,DAI122,DAI111a,DAI111b,Zhao11}. Further analysis shows that the structure transition coincides with the energy evolution of the topology of the constant energy contours(CECs). The main features of the MEs are dominated by the intraorbital particle-hole excitations. Furthermore, it was found that the MEs exhibit a resonance-like peak at $(\pi,\pi)$ with the characteristic energy as high as $0.34$ which is comparable to the energy scale of the collective mode suggested by the ARPES observations\cite{BorisenkoHE}. The structure transitions of the MEs remain in the spin and orbital scenarios of nematic phases, although anisotropy of the MEs develops due to the $C_{4}$ symmetry breaking induced by the nematic orders. In the spin-nematic scenario, the anisotropy of the MEs persists up to the high energy region. While, for the orbital-nematic scenario, it decreases dramatically with the increase of energy and is negligible in the high energy region.

\section{The model and formulas}
The Hamiltonian we use to carry out the calculations can be divided
into three parts: the tight binding part, the Coulomb interacting part and
the part that modeling the spin or orbital driven nematic phase.

We adopt the five orbital tight binding Hamiltonian of
Ref~\cite{GRASER} which reproduces the LDA energy bands. Following the previous studies\cite{Kreisel2015,bandrenm}, a band renormalization factor $z=3$ is introduced. This magnitude of $z$ is compatible with the experimental observations and the numerical simulations\cite{REZa,REZb,FERBER,AICHHORN,YIN,WERNER}. In this way, the tight
binding Hamiltonian reads $H_{0}=\sum_{k,a,b,\sigma}
z^{-1}\epsilon_{ab}(k)C^{+}_{ka\sigma}C_{kb\sigma}$, where $a,b$ and
$\sigma$ are the orbital and spin indices, respectively. The parameters for $\epsilon_{ab}(k)$ are given in
Ref~\cite{GRASER} and the energy unit $eV$ will be used throughout
the paper.

As usual, the Coulomb interaction Hamiltonian $H_{int}$ is given by,
\begin{eqnarray}
H_{int}&=&U\sum_{i,a} n_{ia\uparrow}n_{ia\downarrow}+U^{'}\sum_{i,a<b}n_{ia}n_{ib}+J\sum_{i,a<b}
C^{+}_{ia\sigma}C^{+}_{ib\sigma^{'}}C_{ia\sigma^{'}}C_{ib\sigma} \nonumber\\
&+&J^{'}\sum_{i,a\neq b} C^{+}_{ia\uparrow}C^{+}_{ia\downarrow}
C_{ib\downarrow}C_{ib\uparrow}.
\label{eqn.1}
\end{eqnarray}
Where $n_{ia}=n_{ia\uparrow}+n_{ia\downarrow}$ is the electron number of orbital $a$ at site $i$ with $n_{ia\sigma}=C_{ia\sigma}^{+}C_{ia\sigma}$. $U,U',J,J'$ are the
coefficients of the intraorbital interaction, interorbital
interaction, Hund coupling, and pair hopping terms, respectively.
$U=U^{'}+J+J^{'}$ and $J=J^{'}$ are assumed as required by the
spatial rotational symmetry.

In iron-pnictides, the nematic order breaks the $C_{4}$ symmetry but restores the spin rotational
symmetry. As proposed theoretically, the simple interacting term $H_{sn}=V_{n}\sum_{iab}(\vec{S}_{i,a}\cdot\vec{S}_{i+\delta_{x},b}-
\vec{S}_{i,a}\cdot\vec{S}_{i+\delta_{y},b})$ is adopted to model the spin driven nematicity\cite{CFANG,CKXU,FERNANDES}. Where $V_{n}$ is the spin-nematic order,
$\vec{S}_{i,a}$ is the spin operator of orbital $a$ at site $i$. $\delta_{x}$ and $\delta_{y}$ are unit vectors along the
$x$ and $y$ axes, respectively. Generally, there are other possible spin-nematic terms allowed by symmetry. For simplicity, we focus on the present form of the spin-nematic order. We believe that the main physics of the spin driven nematicity have been captured by the model shown above. While in the orbital-nematic scenario,
the orbital orders are used to model the orbital driven nematicity\cite{ONARI,OOCHEN,OOLV,Zhang}. The explicit form of the orbital orders is established in Ref.~\cite{Zhang} based on the symmetry analysis. For simplicity, we focus on the orbital order of the form
$H_{on}=\sum_{k\sigma}\lambda_{0}(C^{+}_{k\sigma,xz}C_{k\sigma,xz}-C^{+}_{k\sigma,yz}C_{k\sigma,yz})+\lambda_{1}(\cos
k_{x}-\cos k_{y})(C^{+}_{k\sigma,xz}C_{k\sigma,xz}+C^{+}_{k\sigma,yz}C_{k\sigma,yz})$ which dominates the band splitting behaviors around the Fermi level\cite{FESED,FESEL}, where $\lambda_{0}$ and $\lambda_{1}$ are the orbital order parameters. Thus, the full Hamiltonian reads $H_{s}=H_{0}+H_{int}+H_{sn}$
for the spin-nematic phase, while it reads $H_{r}=H_{0}+H_{on}+H_{int}$ for the orbital-nematic phase. With these Hamiltonian, the spin susceptibility is calculated within the random phase approximation(RPA) method. For the spin and orbital scenarios, the quadratic terms $H_{0}$ and $H_{0}+H_{on}$ are used to construct the Green's functions, while the quartic terms
$H_{int}+H_{sn}$ and $H_{int}$ are considered as the interaction between the particle-hole pairs, respectively.

In the present model, the Green's function can be defined in the orbital basis as $\hat{G}_{ab\sigma}(k,\tau)=-\langle T[C_{k a\sigma}(\tau)C_{k b\sigma}^{+}(0)]\rangle$.
Due to the spin rotational symmetry in the present model, it is adequate to study the transverse spin susceptibility which is defined as
$\hat{\chi}_{ab,cd}(q,\tau)=\langle T[S^{-}_{ab}(q,\tau)S_{cd}^{+}(q,0)]\rangle$, where $S^{-}_{ab}(q)=\frac{1}{\sqrt{N}}\sum_{k}C_{a\downarrow}^{+}(k)C_{b\uparrow}(k+q)$ and $S^{+}_{ab}(q)=\frac{1}{\sqrt{N}}\sum_{k}C_{a\uparrow}^{+}(k+q)C_{b\downarrow}(k)$ are the spin flip operators with $N$ the number of the lattice sites.
The bare spin susceptibility can be expressed as  $\hat{\chi}^{0}_{ab,cd}(q,\tau)=-\frac{1}{N}\sum_{k}G_{da\downarrow}(k,-\tau)G_{bc\uparrow}(k+q,\tau)$.
After the fourier transformation, we get $\hat{\chi}^{0}_{ab,cd}(q,i\omega_{n})=-\frac{1}{N\beta}\sum_{k,i\omega_{m}}G_{da\downarrow}(k,i\omega_{m})G_{bc\uparrow}(k+q,i\omega_{m}+i\omega_{n})$,
where $\beta=\frac{1}{T}$ is the inverse of temperature $T$.

In the spin-nematic scenario, the nonzero elements of the interaction matrix read as $\hat{U}_{aa,aa}(q)=U-V_{n}(\cos q_{x}-\cos q_{y})$, $\hat{U}_{aa,bb(a\neq b)}(q)=
J-V_{n}(\cos q_{x}-\cos q_{y})$, $\hat{U}_{ab,ba(a\neq b)}(q)=U^{'}$ and $\hat{U}_{ab,ab(a\neq b)}(q)=J^{'}$. While in the orbital scenario, they read $\hat{U}_{aa,aa}(q)=U$, $\hat{U}_{aa,bb(a\neq b)}(q)=J$, $\hat{U}_{ab,ba(a\neq b)}(q)=U^{'}$ and $\hat{U}_{ab,ab(a\neq b)}(q)=J^{'}$. Thus, the RPA spin susceptibility can be constructed as $\hat{\chi}(q,i\omega_{n})=\hat{\chi}^{0}(q,i\omega_{n})(\hat{I}-\hat{U}(q)\hat{\chi}^{0}(q,i\omega_{n}))^{-1}$, where $\hat{\chi}^{0}$, $\hat{U}$ and $\hat{\chi}$ are the bare spin susceptibility, the interaction matrix and the RPA renormalized spin susceptibility, respectively. The spectra of the MEs can be obtained through the analysis of the imaginary part of the RPA spin susceptibility which reads $\chi^{''}(q,\omega)=\sum_{a,b}Im\hat{\chi}_{aa,bb}(q,\omega)$.

Without loss of generality,
$U=0.35$ and $J=U/4$ is used throughout the paper. The temperature is set to be $T=0.001$. $V_{n}$ and
$\lambda_{i}$ are the order parameters for the spin and orbital scenarios of nematicity, respectively.

\section{The magnetic excitations}
\subsection{The normal state magnetic excitations}
\begin{figure}
\centering\includegraphics[width=0.5\textwidth]{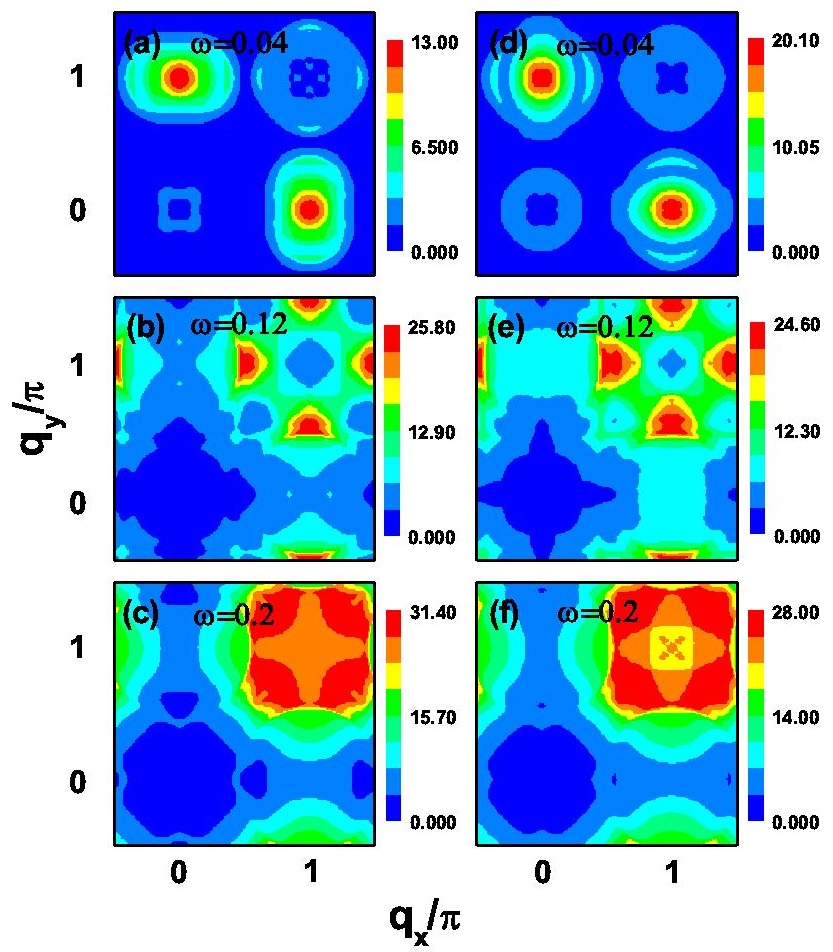}
\caption{(Color online) Energy evolution of the normal state magnetic excitations. (a), (b) and (c) are MEs for $n=6.03$. (d), (e) and (f) are MEs for $n=5.95$.}
\label{f.1}
\end{figure}
We begin with the study of the normal state MEs. In Fig.~\ref{f.1}(a)-(c) and (d)-(f), we show the energy evolution of the MEs for $n=6.03$ and $n=5.95$, respectively. As shown in Fig.~\ref{f.1}(a) for $\omega=0.04$, the MEs are dominated by two commensurate peaks centered at $(\pi,0)$ and $(0,\pi)$. With the increase of energy, the peaks move toward the $(\pi,\pi)$ region. As shown in Fig.~\ref{f.1}(b), incommensurate peaks develop at $(\pi\pm\delta,\pi)$ and $(\pi,\pi\pm\delta)$ for $\omega=0.12$, where $\delta$ denote the incommensurability of the peaks. This gives rise to a diamond-like structure centered at $(\pi,\pi)$. When the energy increases to be $\omega=0.2$, the MEs are dominated by peaks centered at $(\pi\pm\delta,\pi\pm\delta)$ and $(\pi\pm\delta,\pi\mp\delta)$. As shown in Fig.~\ref{f.1}(c), the four peaks of the MEs exhibit a square-like structure. Thus, a commensurability to diamond and to square structure transition occurs for the normal state MEs with the increase of energy. Further study shows that this structure transition persists in a wide doping region. For illustration, we show the energy evolution behavior of the MEs in Fig.~\ref{f.1}(d)-(f) for $n=5.95$. It can be seen that similar structure transition occurs with the increase of energy. Actually, the main features of the MEs for $n=5.95$ are nearly the same with those for $n=6.03$. Thus, the structure transition is a common feature of the normal state MEs. The above established energy evolution behaviors of the MEs are qualitatively consistent with the experimental observations\cite{DAI122a,DAI122,DAI111a,DAI111b,Zhao11}. The structure transition of the MEs persists in a wide doping region. This is also consistent with the INS results\cite{DAI122a,DAI122,DAI111a}.

\begin{figure}
\centering\includegraphics[width=0.7\textwidth]{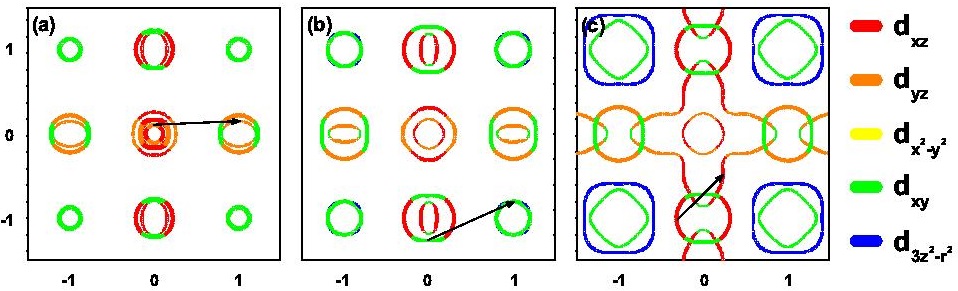}
\caption{(Color online) Schematic shown of the particle-hole scattering between the negative and positive constant energy contours(CECs). The light and deep colors show the dominate orbital components of the CECs of the negative and positive energies, respectively. (a) is the CECs corresponding to $\omega=-0.02$ and $0.02$. (b) and (c) are the same with (a) but for $\omega=-0.04$, $0.08$ and $\omega=-0.1$, $0.1$, respectively.}
\label{f.cec}
\end{figure}

The above mentioned structure transition of the MEs can be qualitatively understood through the analysis of the particle-hole scatterings. In Fig.~\ref{f.cec}, we show the typical particle-hole scattering process between the negative and positive constant energy contours(CECs). In Fig.~\ref{f.cec}(a), the CECs of $\omega=-0.02$ and $0.02$ are shown which correspond to the MEs for $\omega=0.04$. The dominating orbital components of the CECs are shown with colors depicted in Fig.~\ref{f.cec}. For $\omega=-0.02$, the CECs are composed of the two hole pockets around $(0,0)$ marked by light colors, the inner electron pocket around $(\pi,0)$ and $(0,\pi)$ and the hole pockets around $(\pi,\pi)$. While the CECs of $\omega=0.02$ are composed of the outer electron pockets around $(\pi,0)$ and $(0,\pi)$ and two hole pockets around $(0,0)$ marked by deep colors. It should be noticed that the inner hole pocket of $\omega=-0.02$ coincides basically with the outer hole pocket of $\omega=0.02$. Thus, only three hole pockets are visible in Fig.~\ref{f.cec}(a). Due to the nesting effect, the MEs are dominated by the particle-hole scatterings between the hole pockets around $(0,0)$ and the electron pockets around $(0,\pi)$ or $(\pi,0)$. This leads to the commensurate MEs with peaks at $(\pi,0)$ and $(0,\pi)$ for $\omega=0.04$. For $\omega=0.12$, we show the representative particle-hole scatterings between the CECs of $\omega=-0.04$ and $0.08$ in Fig.~\ref{f.cec}(b). The CECs of $\omega=0.08$ are the larger electron pockets around $(\pi,0)$ and $(0,\pi)$, while the remaining parts of the CECs correspond to $\omega=-0.04$. Clearly, there is mismatch between the hole pockets around $(0,0)$ and the electron pockets around $(\pi,0)$ or $(0,\pi)$. In contrast, the electron pocket around $(0,\pi)$ and the hole pocket around $(\pi,\pi)$ are well nested. As a result, the MEs are dominated by the peaks at $(\pi\pm\delta,\pi)$ and $(\pi,\pi\pm\delta)$ which correspond to the scattering process indicated by the black arrow in Fig.~\ref{f.cec}(b). It should be noticed that the above scattering process is intraorbital dominated by $d_{xy}$. When the energy increases to be $0.2$, the corresponding CECs change dramatically in structure. The typical CECs are shown in Fig.~\ref{f.cec}(c) for $\omega=-0.1$ and $0.1$. The CECs for $\omega=0.1$ are the electron pockets centered at $(0,\pi)$ and $(\pi,0)$, while the remaining parts correspond to $\omega=-0.1$. The dominating scattering process is indicated by the black arrow in Fig.~\ref{f.cec}(c) which connects the larger hole pocket centered at $(0,0)$ and the electron pockets around $(0,\pi)$. This gives rise to the square-like MEs with the peaks centered at $(\pi\pm\delta,\pi\pm\delta)$ and $(\pi\pm\delta,\pi\mp\delta)$. Depending on the location of the peaks, the MEs are dominated by the intraorbital scatterings of $d_{xz}$ or $d_{yz}$. From the above analysis, it can be seen that the main features of the MEs are tightly related to the intraorbital particle-hole scatterings. The structure transition occurs due to the fact that the structure of the CECs changes with the increase of energy.

\begin{figure}
\centering\includegraphics[width=0.7\textwidth]{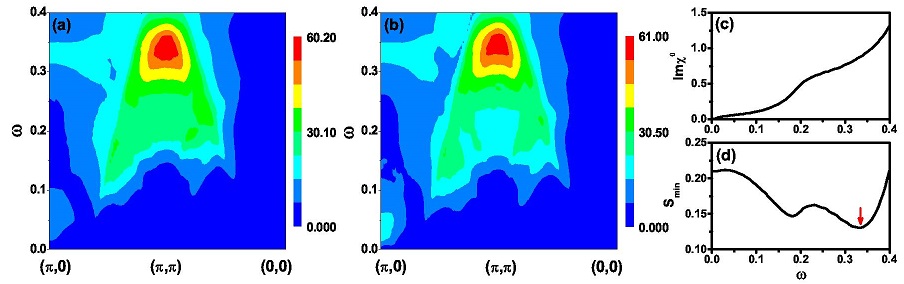}
\caption{(Color online) (a) and (b) are the normal state MEs for $n=6.03$ and $n=5.95$, respectively.
(c) and (d) are the energy dependence of the imaginary part of the bare spin susceptibility and the minimum singular value of $\hat{I}-\hat{U}(q)\hat{\chi}^{0}(q,\omega)$ for $n=6.03$, respectively.}
\label{f.2}
\end{figure}

To show the momentum and energy dependence of the MEs, we plot in Fig.~\ref{f.2} the MEs along the high symmetry direction of the Brillouin zone. Fig.~\ref{f.2}(a) and (b) are the normal state MEs for $n=6.03$ and $n=5.95$, respectively. It can be seen that the main features of the MEs do not change qualitatively with the changing of the doping level. The MEs are composed of two parts in energy-momentum space with the low energy part locating around $(\pi,0)$ and $(0,\pi)$, while the high energy part around $(\pi,\pi)$. In the low energy region of $0<\omega<0.06$, the MEs are dominated by the commensurate peaks around $(\pi,0)$ and $(0,\pi)$. With the increase of energy, significant intensity of the MEs develops around $(\pi,\pi)$. When $0.1<\omega<0.15$, the MEs are dominated by the peaks orienting along the $(\pi,0)-(\pi,\pi)$ direction, giving rise to the above mentioned diamond-like structure of the MEs. With the further increase of energy, intensity peaks develop gradually along the $(\pi,\pi)-(0,0)$ direction. As a result, a diamond to square structure transition occurs in this energy region. Besides, we find that a resonance-like peak develops around $(\pi,\pi)$ with the characteristic energy of $0.34eV$. It is interesting to notice that the recent ARPES data indicates that the band renormalization in the high energy region is caused by the coupling of electrons with a collective mode with characteristic energy of $0.5eV$\cite{BorisenkoHE}. As shown in Fig.~\ref{f.2}(c), the imaginary part of the bare spin susceptibility exhibits no singular behavior around $\omega=0.34$. Thus, the resonance-like behavior of the MEs is attributed to the RPA renormalization of the bare spin susceptibility. We perform the singular value decomposition of $\hat{I}-\hat{U}(q)\hat{\chi}^{0}(q,\omega)$ which is the denominator of the RPA spin susceptibility. The energy dependence of the minimum singular value $S_{min}$ is shown in Fig.~\ref{f.2}(d). It can be seen that $S_{min}$ exhibits minima exactly at $0.34$, thus the resonance-like mode arises from the pole-like effects of the RPA spin susceptibility. Unlike the bound states of the MEs which occur below $2\Delta_{sc}$ in the superconducting state\cite{FESPINa,FESPINb}, the resonance-like mode develops at much higher energies and it is strongly damped due to its deep in the particle-hole continuum. Actually, no real pole occurs for the RPA spin susceptibility due to the finite value of the imaginary part of the bare spin susceptibility as shown in Fig.~\ref{f.2}(c).

\subsection{The MEs in the spin and orbital scenarios of nematicity}
\begin{figure}
\centering\includegraphics[width=0.6\textwidth]{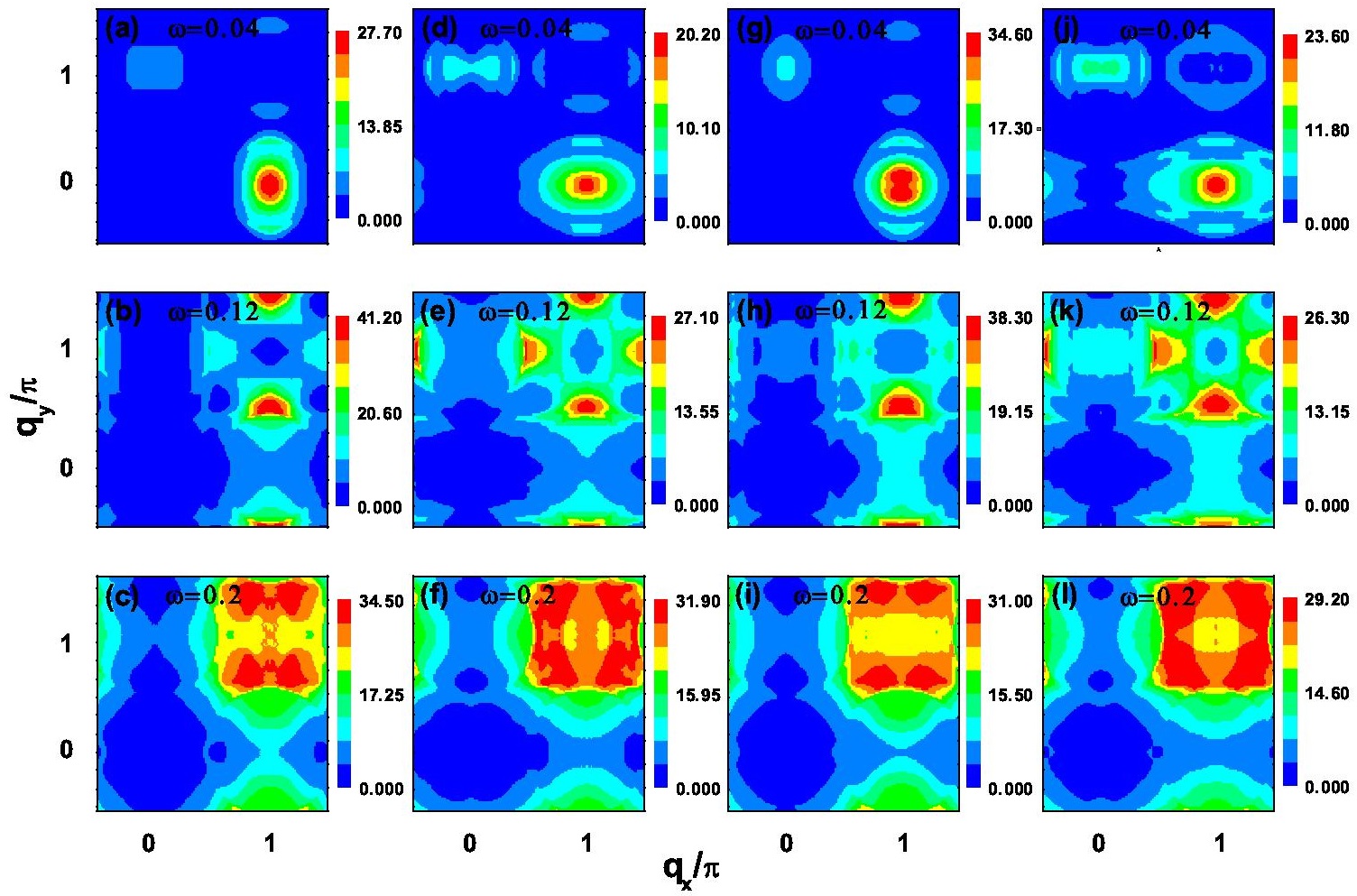}
\caption{(Color online) Energy evolution of the magnetic excitations in the spin and orbital scenarios of nematic phases. (a), (b), (c) and (d), (e), (f) show the energy evolution of the MEs for the spin and orbital driven nematicity for $n=6.03$, respectively.
(g), (h), (i) and (j), (k), (l) show the energy evolution of the MEs for the spin and orbital driven nematicity for $n=5.95$, respectively.}
\label{f.4}
\end{figure}

As revealed by previous experiments\cite{DAIprl2013,DAIscience2014,FEOOa,FEOOb,FEOOc}, the nematic phase is accompanied by the anisotropic magnetic excitations and band splitting between $d_{xz}$ and $d_{yz}$. It was theoretically proposed that the nematicity may originate from the spin or orbital fluctuations\cite{NMREV,CFANG,CKXU,FERNANDES,ONARI}. Here, we study the feedback effects of the spin and orbital scenarios of nematic orders on the behaviors of the MEs. In Fig.~\ref{f.4}, we show the energy evolution of the MEs in the spin and orbital driven nematic phases for $V_{n}=0.01$ and $\lambda_{0}=-0.03$, respectively. In panels (a), (b), and (c) of Fig.~\ref{f.4}, we show the energy evolution of the MEs for the spin case of $n=6.03$. For a low energy of $\omega=0.04$, the MEs are dominated by a commensurate peak centered at $(\pi,0)$. In contrast, the intensity of the MEs around $(0,\pi)$ is much weaker. This is attributed to the spin-nematic orders which break the $C_{4}$ rotational symmetry. For positive $V_{n}$, the spin interaction is enhanced around $(\pi,0)$, while it is depressed around $(0,\pi)$. Thus, the MEs show larger intensity around $(\pi,0)$. When the energy increases to be $\omega=0.12$, the MEs turn to be dominated by the structure around $(\pi,\pi)$. Two peaks centered at $(\pi,\pi\pm\delta)$ dominate the main features of the MEs. There are two additional peaks centered at $(\pi\pm\delta,\pi)$, however they are much weaker in intensity. This is attributed to the spin-nematic order which enhances the spin interaction along the vertical direction while depresses it along the horizontal direction around the $(\pi,\pi)$ region. The four peaks exhibit the similar diamond-like structure with those shown in Fig.~\ref{f.1}(b). With the further increase of energy, the pattern of the MEs changes gradually and it is dominated by four diagonal peaks which form a square-like structure when the energy is around $0.2$. As shown in Fig.~\ref{f.4}(a), (b), and (c), the structure transition of the MEs in the spin-nematic scenario is quite similar to that depicted in Fig.~\ref{f.1}(a), (b), and (c). However, due to the spin-nematicity induced $C_{4}$ rotational symmetry breaking, the MEs is distorted in structure. As shown in Fig.~\ref{f.4}(g), (h), and (i), similar behaviors of the MEs exist for the hole doped case of $n=5.95$. It was checked that the above established features of the MEs persist in the spin-nematic phase when $n$ is around $6.0$.
\begin{figure}
\centering\includegraphics[width=0.6\textwidth]{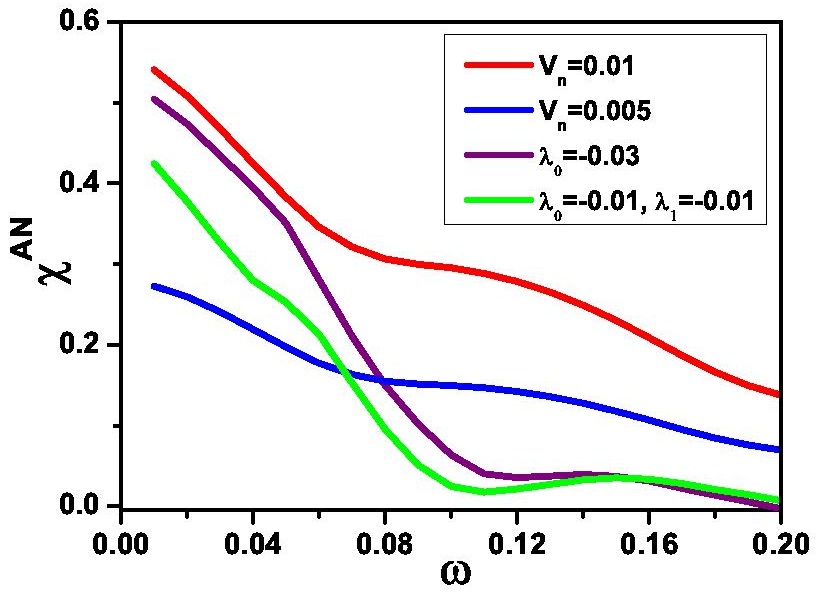}
\caption{(Color online) Energy dependence of the anisotropy of MEs in the spin and orbital scenarios of nematic phases for $n=6.03$.}
\label{f.5}
\end{figure}

It should be noticed that quite similar structure transition of the MEs occurs in the orbital-nematic scenario. In Fig.~\ref{f.4}(d)-(f) and (j)-(l), the energy evolution behaviors of the MEs are shown for $n=6.03$ and $n=5.95$, respectively. A commensurability to diamond structure transition occurs when the energy increases from $\omega=0.04$ to $\omega=0.12$. At a high energy of $\omega=0.2$, the MEs exhibit a square-like structure around $(\pi,\pi)$. It was checked that the structure transition is qualitatively unchanged when the momentum dependent orbital order $\lambda_{1}$ is taken into account. Thus, the established behaviors of the MEs are robust phenomenon in the orbital-nematic case. Compared to the spin-nematic case, the anisotropy of the MEs is weaker in the orbital-nematic scenario. Especially, it weakens significantly with the increase of energy. As shown in Fig.~\ref{f.4}(e) and (k) for the orbital case of $n=6.03$, the MEs are dominated by four peaks with comparable intensity around $(\pi,\pi)$. In contrast, for the spin-nematic case, the vertical peaks have significantly larger intensity than the horizontal ones. This has been shown in Fig.~\ref{f.4}(b) and (h). Considering that the magnitude of $V_{n}$ is smaller than that of $\lambda_{0}$, it is clearly that the spin-nematic order has more prominent influences on the behaviors of the MEs.
\begin{figure}
\centering\includegraphics[width=0.5\textwidth]{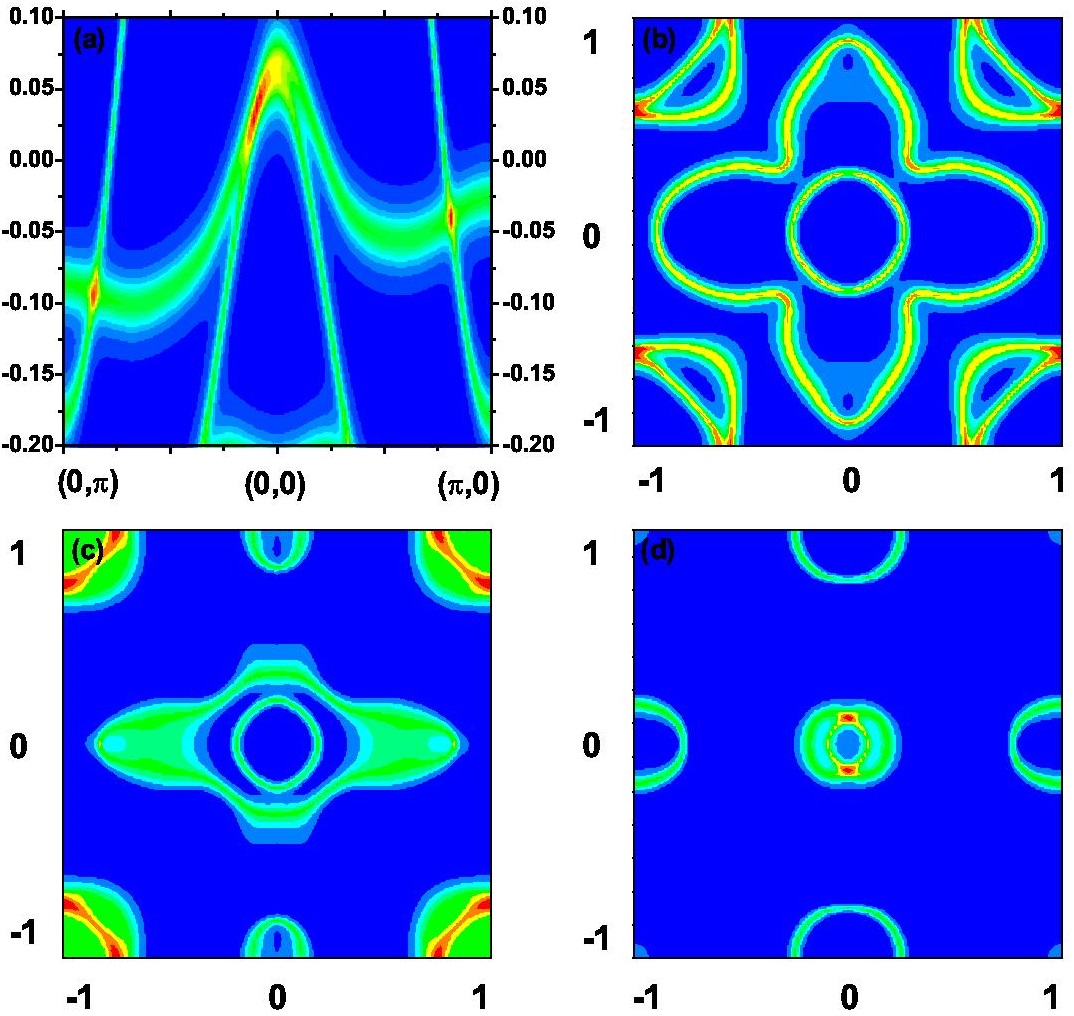}
\caption{(Color online) (a) is the plot of the quasiparticle spectral function in the orbital-nematic state for $n=6.03$, $\lambda_{0}=-0.01$ and
 $\lambda_{1}=-0.01$. (b), (c), and (d) are the plots of the spectral function for $\omega=-0.13$, $-0.06$ and $0.02$, respectively.}
\label{f.6}
\end{figure}

As shown above, both the spin and orbital nematic orders give rise to the anisotropic behaviors of the MEs. However, the energy evolution behaviors of the anisotropy are distinct between these two cases. Here, we introduce a quality $\chi^{AN}(\omega)=\frac{\chi_{a}(\omega)-\chi_{b}(\omega)}{\chi_{a}(\omega)+\chi_{b}(\omega)}$ to characterize the energy dependence of the anisotropy of the MEs, where $\chi_{a}(\omega)=\sum_{|q_{x}|>|q_{y}|}\chi^{''}(q,\omega)$ and $\chi_{b}(\omega)=\sum_{|q_{x}|<|q_{y}|}\chi^{''}(q,\omega)$. In Fig.~\ref{f.5}, the red and blue lines are the energy dependence of $\chi^{AN}(\omega)$ for $V_{n}=0.01$ and $0.005$ for the spin-nematic scenario of $n=6.03$, respectively. $\chi^{AN}(\omega)$ reduces gradually with the increase of energy, and persists to be finite around $0.2$. Thus, the anisotropy of the MEs persists up to the high energy region for the spin-nematic scenario. As shown in Fig.~\ref{f.5}, $\chi^{AN}(\omega)$ decreases with the reduction of $V_{n}$. However, it exhibits quite similar energy dependence when $V_{n}$ varies. The purple and green lines are the energy dependence of $\chi^{AN}(\omega)$ for $\lambda_{0}=-0.03, \lambda_{1}=0$ and $\lambda_{0}=-0.01, \lambda_{1}=-0.01$ for the orbital-nematic scenario, respectively. These two sets of orbital orders give rise to the same splitting energy between the $d_{yz}$ band around $(\pi,0)$ and the $d_{xz}$ band around $(0,\pi)$. The magnitude of the splitting energy is comparable to the experimental data on FeSe\cite{FESED,FESEL}. In this case, $\chi^{AN}(\omega)$ decreases dramatically in the low energy region and it is rather small when the energy is above $0.1$. Thus, the anisotropic MEs are expected to be observed in the low energy region for the orbital-nematic scenario.

The distinct energy evolution behaviors of $\chi^{AN}(\omega)$ between the spin and orbital cases can be attributed to the different origins of the nematic orders. For the spin scenario, the anisotropic MEs arises from the anisotropic interaction induced by the spin-nematic order. Thus, the MEs exhibit anisotropic behaviors up to the high energy region within the RPA treatment of the spin susceptibility. In contrast, for the orbital scenario, the anisotropy of the MEs is directly related to the anisotropic band structure induced by the orbital orders. We find that the band distortions induced by the orbital orders do not occur in the full energy region. As shown in Fig.~\ref{f.6}(a), the anisotropy of the energy bands is mainly reflected in the energy region ranging from the bottom of the $d_{xz}$ band to energies slightly above the Fermi level. For illumination, we show the quasiparticle spectral functions(QSFs) in Fig.~\ref{f.6}(b), (c) and (d) for $\omega=-0.13$, $-0.06$ and $0.02$, respectively. As shown in Fig.~\ref{f.6}(b) for the QSFs near the bottom of the $d_{xz}$ band, there are small anisotropy between the $(0,\pi)$ and $(\pi,0)$ regions. With the increase of energy, significant anisotropy develops in the momentum space. The typical QSFs are shown in Fig.~\ref{f.6}(c) for $\omega=-0.06$. The anisotropy of the QSFs weakens gradually when the energy is above the Fermi level. The typical features of the QSFs are shown in Fig.~\ref{f.6}(d) for $\omega=0.02$. It is interesting to notice that the anisotropy of the QSFs is much stronger when the energy is around the Dirac-cone dispersions locating around $(\pi,0)$ and $(0,\pi)$. As shown in Fig.~\ref{f.6}(a), the Dirac-cone dispersions dominate the energy bands in the energy region from $-0.12eV$ to $-0.01eV$. This energy separation is slightly larger than $0.1eV$ above which the anisotropy of the MEs is rather small in the orbital-nematic case. Thus, the anisotropic MEs are attributed to the Dirac-cone dispersions induced by the orbital orders in the orbital scenario of nematicity. This coincides with the previous study that the Dirac-cone dispersions play the major role in the features of the quasiparticle scattering interference patterns\cite{Zhang}. Thus, the distinct energy evolution behaviors of $\chi^{AN}(\omega)$ can be traced back to the different origins between the spin and orbital scenarios of nematic orders which give rise to anisotropic interaction and distorted band structure, respectively.

\section{Summary and conclusions}
In conclusion, we have studied the behaviors of MEs in the normal and nematic phases of iron-pnictides. In the normal state, the MEs exhibit structure transitions with
the increase of energy. In the low energy region, the MEs are dominated by the commensurate peaks centered at $(\pi,0)$ and $(0,\pi)$. In the intermediate energy, the MEs turn to be dominated by peaks at $(\pi\pm\delta,\pi)$ and $(\pi,\pi\pm\delta)$ which form the diamond-like structure. With the further increase of energy, a square-like structure of the MEs develops in the high energy region. These behaviors of the MEs are attributed to the energy evolution of the CECs. The main features of the MEs are dominated by the intraorbital particle-hole scatterings. Meanwhile, the established energy evolution behaviors of the MEs are qualitatively consistent with the INS observations\cite{DAI122a,DAI122,DAI111a,DAI111b,Zhao11}. Besides, we find that the MEs exhibit a resonance-like peak around $(\pi,\pi)$ with the characteristic energy as high as $0.34$ which is comparable to that suggested by the ARPES data\cite{BorisenkoHE}. This mode shares the similar origin with the magnetic resonance mode in the superconducting state\cite{FESPINa,FESPINb}, although it is strongly damped due to its deep in the particle-hole continuum.

The structure transition of the MEs remains when the spin and orbital nematic orders are taken into account, although anisotropy develops due to the nematicity induced $C_{4}$ symmetry breaking. However, the MEs exhibit distinct energy evolution between the spin and orbital scenarios of nematicity. The anisotropy of the MEs persists to the high energy region for the spin-nematic case. In contrast, it reduces dramatically in the low energy region and is negligible in the high energy region for the orbital-nematic case. These are directly related to the different origins between the spin and orbital scenarios of nematic phases. Thus, these distinct behaviors of the MEs between the spin and orbital cases can be used to distinguish the origin of the nematic phase by using the INS technique.

\section{Acknowledgments}
This work was supported by the National Natural Science Foundation of China (Grants No.~11647072 and No.~11404278).\\

\end{document}